\documentclass[a4paper,11pt]{article}
\pdfoutput=1
\usepackage{jcappub} 
\usepackage[T1]{fontenc} 
\usepackage{slashed}
\usepackage{bm}

\newcommand{\rhoSSB}{\rho_\mathrm{SSB}}
\newcommand{\lcom}{L_\mathrm{c}}
\newcommand{\adw}{a_\mathrm{DW}}
\newcommand{\Mpl}{M_\mathrm{pl}}
\newcommand{\Lpl}{L_\mathrm{pl}}
\DeclareMathOperator{\sech}{sech}

\usepackage{xcolor}
\newcommand{\hl}[1]{\textcolor{magenta}{#1}}
\renewcommand{\hl}[1]{#1}

\title{Dark sector domain walls could explain the observed planes of satellites}

\author{Aneesh P. Naik}
\author{and Clare Burrage}
\affiliation{School of Physics \& Astronomy, University of Nottingham,\\University Park, Nottingham NG7~2RD, UK}
\emailAdd{aneesh.naik@nottingham.ac.uk}

\abstract{The observed `planes of satellites' around the Milky Way and other nearby galaxies are notoriously difficult to explain under the $\Lambda$CDM paradigm. Here, we propose an alternative solution: domain walls arising in theories with symmetry-breaking scalar fields coupled to matter. Because of the matter coupling, satellite galaxies experience fifth forces as they pass through domain walls, leading to a subset of satellites with orbits confined to the domain wall plane. We demonstrate this effect using simple simulations of a toy model comprising point-like satellites and an infinite domain wall, and explore the efficacy of various planarity metrics in detecting this effect. We believe this is the first potential `new physics' explanation for the observed planes of satellites which does not do away with dark matter.}

\begin{document}
\maketitle
\flushbottom

\section{Introduction}
\label{S:Introduction}

The standard model of cosmology, $\Lambda$CDM, has proven to be a fantastically successful paradigm, accounting for a myriad of independent observations on different scales. However, problems begin to emerge when one `zooms in' to small scales, i.e. the regime of individual galaxies and their satellites.

A comprehensive review of these small-scale challenges to $\Lambda$CDM is given by Ref.~\cite{Bullock2017}. Other than the `planes of satellites' studied in the present work, they describe a number of challenges, including the `core-cusp problem': the (indirect) observation of `cored' dark matter (DM) haloes \citep{Moore1994, Flores1994, Walker2011, Oh2015} is possibly in tension with the `cuspy' haloes predicted in $\Lambda$CDM simulations \citep{Navarro1997, Navarro2010}. A second example is the `missing satellites' problem: the number of observed satellite galaxies of the Milky Way \citep{Kim2018, Drlica2020} is significantly smaller than the number one might predict from simulations \citep{Springel2008, Garrison2014, Griffen2016}. Ref.~\cite{Bullock2017} also describes a family of problems under the heading ``Regularity in the Face of Diversity''. This is a series of tight empirical correlations between dynamical properties of galaxies (often principally governed by their dark matter component) and their baryonic properties \citep{McGaugh2012,McGaugh2016}. While such correlations ought to be expected (bigger dark matter haloes should host bigger, brighter galaxies), the tightness of the correlations can be difficult to explain.

Various solutions to these problems have been proposed, often invoking baryonic physics, e.g. supernova feedback or photoionising ultraviolet background radiation. Taking the three small-scale problems listed above for illustration, baryonic physics can solve them by respectively changing the central slope of dark haloes \citep{Pontzen2012, DiCintio2014}, suppressing galaxy formation in smaller dark matter haloes \citep{Efstathiou1992, Shen2014}, and more tightly coupling galaxies with their dark matter halo hosts \citep{Keller2017, Navarro2017}.

Despite these promising results, the implementation of baryonic physics in cosmological simulations is still a relatively young and rapidly developing field \citep{Vogelsberger2020}, and so the efficacies of these mechanisms in solving the small-scale problems is far from certain and subject to ongoing debate. For this reason, many investigations have instead sought alternative solutions in `new physics', i.e. modifications to gravity or more exotic theories of dark matter. Taking the core-cusp problem as an example, with `fuzzy dark matter' \citep{Hu2000}, the bosonic properties of the dark matter particle lead to the formation of halo cores, while under `chameleon' gravity theories \citep{Khoury2004} the presence of a scalar fifth force can lead to a false kinematic inference of a halo core \citep{Lombriser2015, Naik2019}.

In this article, we consider another of these small-scale problems: the `planes of satellites'. Ref.~\cite{Pawlowski2018} provides a comprehensive review of this subject, but to summarise: it has long been observed that the satellite galaxies of the Milky Way \citep{Kunkel1976, Lynden1976, Kroupa2005, Metz2007PoS, Metz2008, Metz2009, Pawlowski2012, Pawlowski2013, Pawlowski2014NewMembers, Pawlowski2015, Pawlowski2016}, Andromeda/M31 \citep{Koch2006, McConnachie2006, Metz2007PoS, Ibata2013, Conn2013}, and other nearby galaxies \citep{Chiboucas2013, Ibata2014SDSSPoS, Ibata2015, Tully2015, Muller2016, Muller2017, Muller2018} orbit their hosts in vast, co-rotating planes. In the case of Andromeda, the satellite population appears bimodal, comprising a subset of satellites closely aligned with the plane and a second, more isotropic subset. The extent to which such systems are to be expected under $\Lambda$CDM is the subject of lively debate. Various analyses of $\Lambda$CDM cosmological simulations have found that systems exhibiting comparable planarity (both spatially and kinematically) arise with frequencies of around 0.1\%--1\% \citep{Pawlowski2014Millennium, Pawlowski2014ELVIS, Muller2018, Ibata2014SimPoS}. However, these findings are not without controversy. For example, some studies have argued that these numbers can be increased with a more careful accounting for the `look-elsewhere' effect \citep{Cautun2015}, or by selecting simulated systems with either an infalling LMC analogue \citep{Samuel2021} or a population of satellites with similar radial clustering to those of the Milky Way \citep{Sawala2022}. Nonetheless, it remains the case that the planes of satellites appear to be unusual systems in $\Lambda$CDM, and it is particularly surprising that they are seen around both major galaxies of the Local Group.

Among the small-scale problems, the satellites planes are unusual (although similar arguments apply to other unresolved problems regarding the phase space distributions of satellites; see Ref.~\cite{Pawlowski2021}) because it is difficult to find an explanation for their origin with baryonic physics: satellites typically inhabit sufficiently large distances from the host galaxy such that the dynamics are largely governed by the dark matter distribution. Some studies have instead argued that group accretion events or a preferred direction of accretion (e.g., along cosmic filaments) could lead to highly anisotropic satellite distributions \citep{Zentner2005, Libeskind2005, Lovell2011, Libeskind2011, Libeskind2015, Angus2016, Samuel2021}, but ultimately these effects are already included in cosmological simulations where they seldom give rise to the high degree of observed planarity. Another idea is that the Local Group satellites are tidal dwarf galaxies (TDGs): dwarf galaxies formed from the gravitational collapse of tidal debris resulting from a major galaxy merger \citep{Zwicky1956, Barnes1992, Duc2012}. If several TDGs form from a single tidal tail, they would approximately share the same orbital plane and direction. This has been proffered as a potential origin for the observed planes of satellites \citep{Kroupa2005, Metz2007TDG, Pawlowski2011, Fouquet2012, Pawlowski2012, Hammer2013, Yang2014, Bilek2021}. Indeed, TDGs have been observed to form in external galaxies (e.g.~\cite{Mirabel1992}). However, it appears unlikely that the Local Group satellites are TDGs, for multiple reasons. First, one would expect TDGs (at least those which have been recently formed) to have systematically higher metallicities than primordial dwarf galaxies as the former objects are formed from pre-enriched material \citep{Duc2012}, but this trend is not observed for the Local Group satellites \citep{Kirby2013, Collins2015}. Second, TDGs are expected to be highly baryon-dominated, hosting very little dark matter \citep{Barnes1992}, while the Milky Way satellites have very high measured mass-to-light ratios \citep{McConnachie2012}, although it has been argued that these mass-to-light ratios might be overestimated as a result of disequilibrium effects \citep{Yang2014}.

Given these issues with the conventional explanations for the satellite planes, a search for possible solutions with new physics is clearly warranted. It is perhaps surprising then that there has been very little work to date in this area. One key exception to this is Modified Newtonian Dynamics (MOND;~\cite{Milgrom1983}): satellite planes do not arise naturally in MOND, but the TDG explanation described above is more palatable under MOND, because the anomalously high mass-to-light ratios inferred for the Milky Way satellites is unrelated to their dark matter content (which is zero in MOND), but due to their MOND-enhanced dynamics \citep{Kroupa2012, Bilek2018}.

Beyond MOND, there has been precious little work on the satellite planes problem under alternative cosmological models, e.g. theories which add novel fields or forces while nonetheless retaining dark matter (unlike MOND). The only example of such an investigation we found in the literature was that of Ref.~\cite{Solis2021}, in which it was shown that a bosonic scalar dark matter component in a multi-state configuration can lead to anisotropic satellite distributions. Here, however, the satellites do not inhabit a single plane, but the infinite family of planes containing the major axis of the dark matter halo. Thus, the applicability of this idea to the observed planes of satellites is unclear.

The present work attempts to address this gulf. We investigate the planar clustering of satellite galaxies as a result of domain walls arising in theories with symmetry-breaking scalar fields coupled non-minimally to matter. Such theories have been studied in a number of other cosmological contexts, such as topological defect dark matter theories \citep{Stadnik2020} or symmetron theories \citep{Dehnen1992, Gessner1992, Pietroni2005, Hinterbichler2010, Hinterbichler2011}. We specifically consider the symmetron framework and couch the study in symmetron terminology, but note that the relevant physics and phenomenology---and thus our results---are common to any scalar theory with topological defects and a non-minimal coupling to matter.

In such theories, domain walls arise at the time of symmetry breaking, forming the boundaries between regions occupying different potential minima. As a result of the matter coupling, these domain walls are significantly different from uncoupled domain walls in several key respects. First, they are `pinned' to overdensities and stabilised by them \citep{Llinares2014SymmDW, Pearson2014, Llinares2019}. Second, they are more numerous in the late Universe than classical domain walls, which typically occur just once per horizon volume \citep{Llinares2014SymmDW}. Applying these ideas to cosmic scales, one expects the domain wall network to pin to structures of the cosmic web. This is the behaviour seen in the cosmological symmetron simulations of Refs.~\cite{Llinares2013, Llinares2014Sims}. It is thus not unreasonable, under this theory, to expect a domain wall to pass through the Milky Way and Andromeda, the largest galaxies of the Local Group. Because of the scalar's non-minimal coupling to matter and the large scalar field gradient through the wall, a fifth force is sourced by the wall, coupling to any standard model or dark matter particles within the range of the force. Because of this attractive force, satellite galaxies preferentially occupy the domain wall plane. We explore this idea for the first time in this article.

This work serves primarily as a preliminary `proof of concept'. We simulate the motions of satellite galaxies (here modelled as simple tracer particles) in the presence and absence of a domain wall, observe the resulting orbital behaviours, and propose appropriate observational diagnostics. As we discuss in our concluding remarks (Sec.~\ref{S:Conclusion}), there are many ways in which reality is more complex than this simple setup, and we propose future work to truly ascertain the viability of domain walls as an explanation for the observed planes of satellites.

This work is structured as follows. In the following section, we introduce the symmetron theory and its domain wall phenomenology. In Section~\ref{S:Simulations}, we describe the simulations we employ to investigate satellite distributions under this theory. We present our results in Section~\ref{S:Results}, before some discussion and concluding remarks in Section~\ref{S:Conclusion}. All of our simulation code and plotting scripts are made publicly available at \url{https://github.com/aneeshnaik/PoSDomainWalls/}.

\section{The Symmetron}
\label{S:Symmetron}

First studied by \cite{Hinterbichler2010, Hinterbichler2011}, the symmetron utilises a general scalar-tensor action of the form
\begin{equation}
\label{E:STAction}
\begin{split}
S = & \frac{c^3}{8\pi G}\int d^4x\sqrt{-g}\left[\frac{1}{2}R - \frac{1}{2}\nabla_{\mu}\phi\nabla^{\mu}\phi - V(\phi)\right]\\ & + S_{m}[A^2\left(\phi\right) g_{\mu\nu},\psi_{i}],
\end{split}
\end{equation}
where $g_{\mu\nu}$ is the Einstein frame metric with determinant $g$, $R$ is the Ricci scalar, $\psi_{i}$ represents the various matter fields in the standard model subject to action $S_m$\footnote{As an example, the action for a toy Standard Model, adopting natural units, is \cite{Burrage2018ScaleInvariance} 
\begin{align}
    S_{m} [A^2\left(\phi\right) g_{\mu\nu},\psi_{i}]= \int d^4 x \sqrt{-g} & \left( -\frac{1}{2} A^2(\phi)g^{\mu\nu}\partial_{\mu} h \partial_{\nu}h +\frac{1}{2}A^4(\phi) \mu_H^2 h^2 -\frac{\lambda_H}{4\!}A^4(\phi)h^4 \right. \\
    & \left. -A^2(\phi)\bar{\psi}i\overleftrightarrow{\slashed{\partial}} \psi - yA^4(\phi)\bar{\psi}h\psi \right)\;,
\end{align}
where the real scalar field $h$ plays the role of the  Higgs, and $\psi$ describes a Dirac fermion. $\mu_H$ and $\lambda_H$ parameterise the Higgs potential, and $y$ is the Yukawa coupling.}, and $\phi$ is the novel scalar field, which is dimensionless in this form of the action. The scalar field potential $V(\phi)$ and coupling to matter $A(\phi)$ can then be chosen so as to yield a symmetron screening mechanism. A variety of such functional forms are possible (e.g.~\cite{Brax2012, Burrage2016}), but we confine our investigation to the original 3-parameter formulation with a quartic potential and quadratic coupling \citep{Hinterbichler2010, Hinterbichler2011}:
\begin{align}
    V(\phi) &= \frac{1}{\Lpl^2}\left[-\frac{1}{2}\left(\frac{\mu}{\Mpl}\right)^2 \phi^2 + \frac{1}{4}\lambda\phi^4\right];\label{E:SymmetronPotential}\\
    A(\phi) &= 1 + \frac{1}{2}\left(\frac{\Mpl}{M}\right)^2\phi^2,\label{E:SymmetronCoupling}
\end{align}
where $\Mpl \equiv \sqrt{\hbar c / 8\pi G}$ and $\Lpl \equiv \sqrt{8\pi G \hbar / c^3}$ are the reduced Planck mass and length respectively. The three input parameters for the theory are the two mass scales $M, \mu$ and the dimensionless $\lambda$. Note that the slightly altered forms of Eqs.~(\ref{E:SymmetronPotential}) and (\ref{E:SymmetronCoupling}) compared with those given by Refs.~\cite{Hinterbichler2010, Hinterbichler2011} (in particular the appearance of factors of $\Mpl$ and $\Lpl$) stem from our choice of `physical' units over natural units, i.e. our retention of factors of $c$ and $\hbar$; the physics of the theory is entirely unchanged. Current constraints on the parameters of the symmetron model are summarised in Refs.~\cite{Burrage2018ChameleonTests, Brax2021}, while constraints on the model parameters assuming that the topological defects that constitute (some of) the dark matter density of the universe can be found in Ref.~\cite{Stadnik2020}.

Considering a static scalar field (i.e. vanishing time derivatives), extremising the action (\ref{E:STAction}) with respect to the scalar field then leads to an equation of motion
\begin{equation}
\label{E:SymmEOM}
    \nabla^2 \phi - \left(\frac{\rho}{\rhoSSB} - 1 \right) \frac{\phi}{2\lcom^2} - \frac{\lambda \phi^3}{\Lpl^2} = 0.
\end{equation}
Note that we have assumed $\phi \ll M/\Mpl$; this holds true for the parameters we will go on to use in our simulations. Two new physical quantities have been defined in terms of the original symmetron parameters: 
\begin{align}
    \lcom &= \frac{1}{\sqrt{2}}\frac{\Mpl}{\mu}\Lpl;\\
    \rhoSSB &= \left(\frac{M^2 \mu^2}{\Mpl^4}\right)\frac{\Mpl}{\Lpl^3}.
\end{align}
The Compton wavelength $\lcom$ sets the length scale over which the scalar field responds to a matter distribution, while $\rhoSSB$ is the threshold density for spontaneous symmetry-breaking.

Spontaneous symmetry breaking is the key aspect of the symmetron model. Considering an infinite box of density $\rho$, if $\rho > \rhoSSB$ then the solution to the equation of motion (\ref{E:SymmEOM}) is $\phi=0$. On the other hand, if $\rho < \rhoSSB$ then the scalar field can adopt either of two solutions
\begin{equation}
\label{E:SymmetronPhiMin}
    \phi_\textnormal{min} = \pm v \sqrt{1 - \frac{\rho}{\rhoSSB}},
\end{equation}
where $v$ is the magnitude of the vacuum expectation value (VEV) of the scalar field, given by
\begin{equation}
    v = \frac{1}{\sqrt{\lambda}}\frac{\mu}{\Mpl}.
\end{equation}

Because of the scalar-matter coupling in (\ref{E:STAction}), matter is accelerated under gradients of the scalar field. The acceleration of a test particle due to the scalar field (the so-called `fifth force') is
\begin{equation}
\label{E:FifthForce}
    \bm{a}_5 = - c^2 \left(\frac{\Mpl}{M}\right)^2 \phi \bm{\nabla} \phi.
\end{equation}
Note that in the unbroken symmetry (i.e. $\rho > \rhoSSB$) regime where $\phi=0$, $\bm{a}_5 = 0$. The fifth force is then said to be `screened'. Only unscreened mass can source or couple to the fifth force.

The final aspect of the symmetron model worth mentioning is the phenomenon of domain walls, which is the subject of the present investigation. Equation~\ref{E:SymmetronPhiMin} demonstrates that there are two unscreened solutions $\phi_\mathrm{min}$ for the scalar field, differing only in sign. If two patches of space adopt opposing minima, they will be separated by a domain wall, across which the field passes from one solution to the other via $\phi=0$. The functional form for the scalar field profile across a domain wall \citep{Llinares2014SymmDW} is
\begin{equation}
\label{E:DWProfile}
    \phi_\mathrm{DW}(z) = |\phi_\mathrm{min}|\tanh\left(\frac{|\phi_\mathrm{min}|}{v}\frac{z}{2 \lcom}\right).
\end{equation}
In a cosmological context, when the background matter density $\bar{\rho}_m \ll \rhoSSB$ (so $|\phi_\mathrm{min}| \approx v$), the thickness of the domain wall is the Compton wavelength.

In the absence of any matter coupling, domain walls arise once per particle horizon volume. If this were to be the case here, it would be highly unlikely that a domain wall should happen to pass through the Milky Way. Fortunately, as a consequence of the matter coupling, the characteristic distance $D$ today between symmetron domain walls is not the particle horizon, but is instead related to the typical co-moving distance between the matter under-densities at which the symmetry was first broken. Using a simplified semi-analytic treatment, Ref.~\cite{Llinares2014SymmDW} found that $D$ depends primarily on the redshift of symmetry breaking $z_\mathrm{SSB}$, with $D \sim 1~\mathrm{Mpc}$ for an early symmetry breaking ($z_\mathrm{SSB}=1000$) and $D \sim 800~\mathrm{Mpc}$ for a late symmetry breaking ($z_\mathrm{SSB}=1$). Thus, given a sufficiently early $z_\mathrm{SSB}$, the domain wall distances are significantly smaller than the particle horizon. Given also the fact that these walls are pinned to and stabilised by matter overdensities \citep{Llinares2014SymmDW, Pearson2014, Llinares2019}, it is then not unreasonable to suppose that the Galaxy might host a domain wall.

Ref.~\cite{Llinares2014SymmDW} also gives expressions for the surface energy density and the fraction of the cosmic energy budget stored in symmetron domain walls $\Omega_\mathrm{DW}$, and compute values for $\Omega_\mathrm{DW}$ across cosmic history and across the allowed region of symmetron parameter space (see Fig.~2 in that work). They find $\Omega_\mathrm{DW} < 1$ everywhere, and $\Omega_\mathrm{DW} \ll 1$ unless the symmetry breaking is quite early ($z_\mathrm{SSB} \approx 1000$). There is thus no danger of the domain walls over-closing the Universe.

\section{Simulations}
\label{S:Simulations}

To explore the effect of a symmetron domain wall on a population of satellite galaxies, we run two simulations, one with a domain wall and one without. In these simulations, the satellites are treated simply as massless point particles moving frictionlessly in static external potentials. In other words, we neglect dissipational effects, hydrodynamics, tidal interactions with the host galaxy, mutual interactions between satellites, and time-dependence of the host potential.

In both simulations, the host galaxy is modelled with a static, spherical NFW profile \citep{Navarro1996, Navarro1997} with a virial mass of $10^{12}~M_\odot$ and a virial concentration of 10, approximately resembling the Milky Way (e.g.~\cite{McMillan2017}). 

In the simulation with a domain wall, the domain wall is infinite in extent, and centred around $z=0$. The vertical (i.e. perpendicular to the wall) scalar field profile is given by Eq.~\ref{E:DWProfile} with the additional assumption that $\bar{\rho}_m \ll \rhoSSB$ (so $|\phi_\mathrm{min}| \approx v$). The fifth force due to the domain wall is $\bm{a} \propto \varphi \bm{\nabla} \varphi$ (see Eq.~\ref{E:FifthForce}), so the satellites feel a vertical acceleration
\begin{equation}
\label{E:DWAcc}
    a(z) = - \adw\tanh\left(\frac{z}{2 \lcom}\right)\sech^2\left(\frac{z}{2 \lcom}\right),
\end{equation}
where
\begin{equation}
    \adw \equiv \frac{c^2}{2\lcom} \frac{1}{\lambda}\left(\frac{\mu}{M}\right)^2.
\end{equation}
The two parameters describing the domain wall are thus $\adw$ and $\lcom$, which set the characteristic acceleration and width respectively.  In our domain wall simulation, we set these to typical galactic scales: $\adw = 5 \times 10^{-11}~\mathrm{m / s^2}$ ${(\approx 1.6 ~\mathrm{Mpc/Gyr^2})}$, $\lcom = 10~\mathrm{kpc}$. We show results for this parameter choice, but we have also checked that our results are robust against variations in these parameters; see Sec.~\ref{S:Conclusion}. 

In terms of the fundamental model parameters, our choice of $\adw$ and $\lcom$ corresponds to choosing $\mu c^2 = 5 \times 10^{-28} \mbox{ eV}$ and $\lambda (Mc^2/\mbox{GeV})^2 = 6 \times 10^{-67}$. To ensure that most satellite galaxies are not screened by the fifth force, we require that their average density (taking some approximate numbers: $\rho_\mathrm{sat} \approx 10^{-21}~\mathrm{kg/m^3} \approx 10^{6} \bar{\rho}_m$, based on a mass of $\sim10^{8} M_\odot$ and size of $\sim1~\mathrm{kpc}$) be lower than $\rho_{\rm SSB}$, \hl{implying a lower bound on $M$, $(M/\Mpl)^2 > 10^{-5}$}. Such a choice of parameters means that our scalar field cannot be also connected to an explanation for dark energy, where $\mu M \sim H_0 M_{\rm Pl}$ is typically required \cite{Hinterbichler2010, Hinterbichler2011}, but is compatible with the topological defects making up a fraction of the current dark matter density \cite{Stadnik2020}. Also, this constraint on $\rhoSSB$ translates to the requirement that the scale factor at symmetry breaking $a_\mathrm{SSB} \lesssim (\rho_\mathrm{sat}/\bar{\rho}_m)^{-3}\approx 10^{-2}$, so that $z_\mathrm{SSB} \gtrsim 100$. So, according to Ref.~\cite{Llinares2014SymmDW} (see also Sec.~\ref{S:Symmetron}), the typical domain wall separation in our case is on $10~\mathrm{Mpc}$ scales or smaller. \hl{There is also an upper limit on $M$, stemming from our assumption that the (repulsive) gravitational force due to the energy density stored in the domain wall \citep{Vilenkin1994} has a negligible dynamical imprint on the system, i.e. it is subdominant to the scalar fifth force at $\sim\mathrm{kpc}$ distances and the Milky Way's gravitational attraction at $\sim 100~\mathrm{kpc}$ distances. Both of these requirements are met if $(M/\Mpl)^2 \ll 1$.} 

At the start of the simulations, we randomly sample each satellite-host distance and satellite speed from uniform distributions $[0, 400]~\mathrm{kpc}$ and $[0, 0.98v_\mathrm{esc}]$ respectively, where $v_\mathrm{esc}$ is the escape speed \emph{at the distance of the given satellite}. The angular position of the satellite and the direction of its velocity vector are then uniformly chosen from a spherical surface. In each simulation, we sample 2500 satellites in this way. This is of course an unrealistically large collection of luminous satellites for a Milky Way-like host (cf. $\sim$50 known satellites of the Milky Way \cite{Drlica2020}), but the increased statistical power better equips us to distinguish the imprint of the domain wall when contrasting the two simulations. Note that the satellite distribution being initially isotropic and radially uniform is also unrealistic. In particular, it is likely that the satellite infall typically has a preferred direction, which can have an impact on the subsequent kinematic and spatial clustering of satellites. While this effect on its own has been shown to be insufficient to explain the observed planes of satellites (see Sec.~\ref{S:Introduction}), it will be interesting in the future to consider its impact alongside that of the domain wall.

Given these initial conditions, we evolve the satellites forward in time using a leapfrog integrator, i.e. at each timestep $i$ the velocity $\bm{v}$ and position $\bm{x}$ of each satellite is updated via
\begin{equation}
\begin{split}
\bm{v}^{i + 1/2} &= \bm{v}^{i - 1 / 2} + \bm{a}(\bm{x}^i) \Delta t; \\
\bm{x}^{i + 1} &= \bm{x}^i + \bm{v}^{i + 1/2} \Delta t,
\end{split}
\end{equation}
where $\Delta t$ is the finite timestep size to be chosen manually, and $\bm{a}(\bm{x})$ is the total acceleration, comprising the radial NFW acceleration plus---in the simulation with a domain wall---the vertical acceleration towards the domain wall (Eq.~\ref{E:DWAcc}). We find that a timestep $\Delta t = 10^{12}~\mathrm{s}$ $(\approx 3 \times 10^{-5}~\mathrm{Gyr})$ gives accurate, converged results. The positions and velocities are `desynchronized' in the sense that $\bm{x}$ is evaluated at integer timesteps, while $\bm{v}$ is evaluated halfway between timesteps, hence the half-integer superscripts above. The initial desynchronization is achieved via
\begin{equation}
\begin{split}
    \bm{v}^{-1/2} &= \bm{v}^0 - \bm{a}(\bm{x}^0)\frac{\Delta t}{2},
\end{split}
\end{equation}
where quantities with the superscript 0 are evaluated at the initial time.

We run each simulation for $3 \times 10^{18}~\mathrm{s}$ $(\approx 95~\mathrm{Gyr})$. This is many times longer than the age of the real Universe, but once again we obtain a greater statistical power through the ability to take simulation snapshots over many dynamical times.

\begin{figure*}
    \centering
    \includegraphics{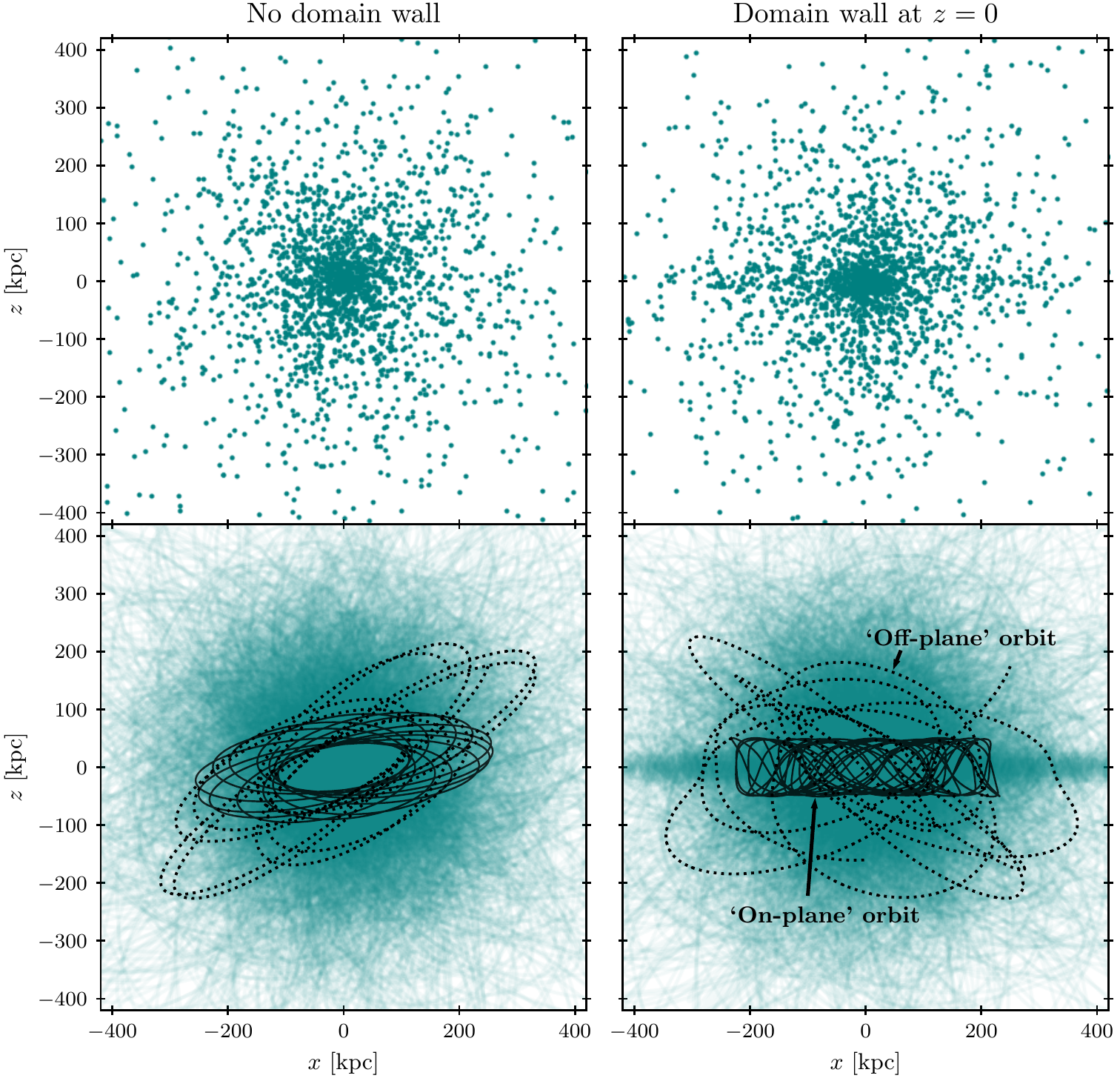}
    \caption{\textit{Upper:} Instantaneous positions of 2500 satellites at the end of simulations with (right) and without (left) a dark domain wall fixed at $z=0$. \textit{Lower:} Each faint line depicts an orbital trajectory of a satellite throughout the $\sim$100~Gyr simulation. The trajectories of a random subsample of 500 satellites are shown here. Because of the domain wall's fifth force, there is a distinct subset of `on-plane' satellites confined to a thin $\sim$10~kpc plane around $z=0$. The two black trajectories in the lower right panel highlight the orbits of two randomly chosen satellites: one on-plane (solid) and off-plane (dotted). The highlighted trajectories in the lower left panel represent the same two satellites, now in the absence of the domain wall.}
    \label{F:SatPositions}
\end{figure*}

\begin{figure*}
    \centering
    \includegraphics{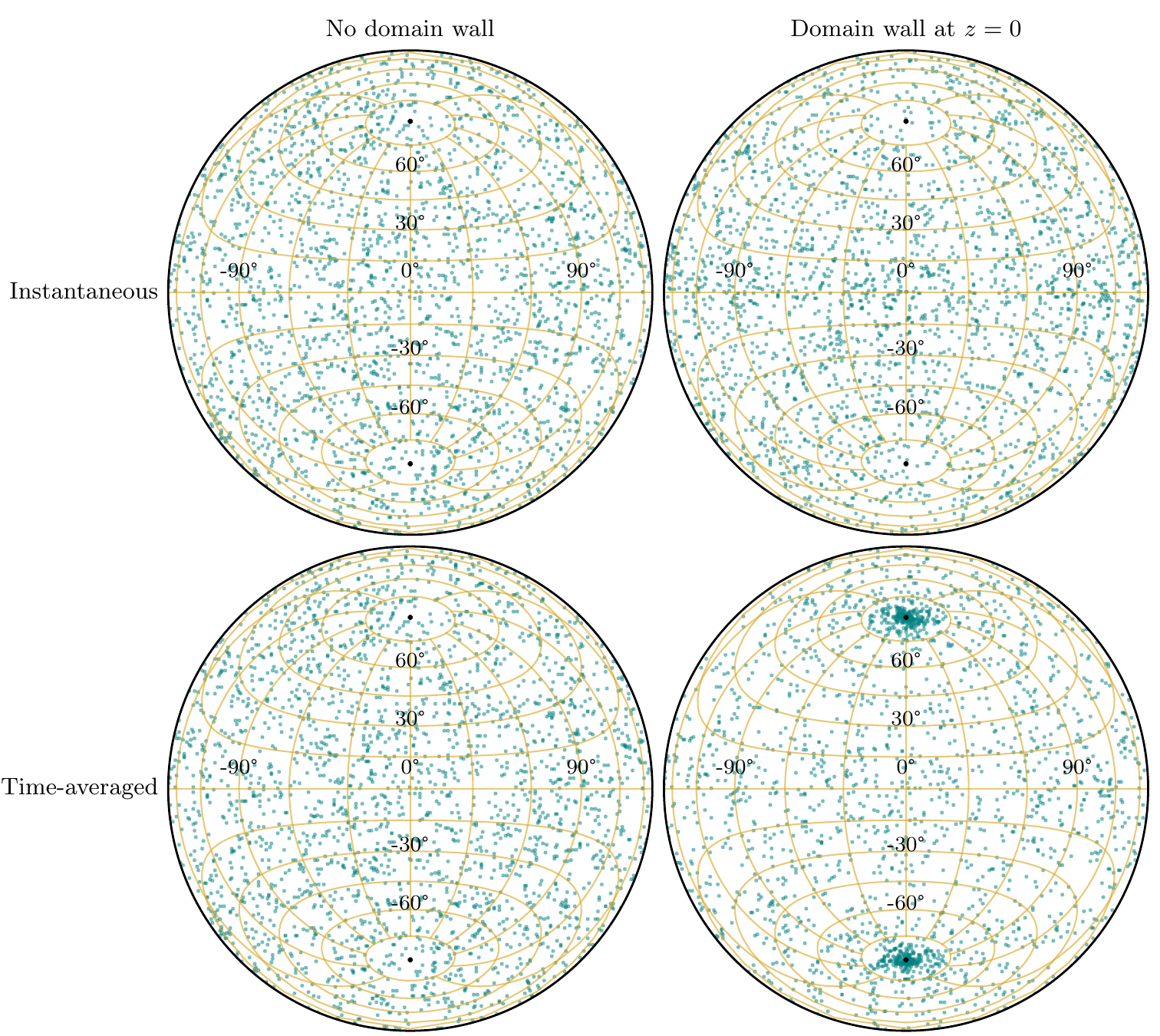}
    \caption{Satellite orbital poles in the simulation with (right) and without (left) a domain wall, both instantaneously at the end of the simulation (upper) and time-averaged over the simulation (lower). While the domain wall does not visibly cause an instantaneous clustering of orbital poles, there is clear clustering on average.}
    \label{F:OrbitalPoles}
\end{figure*}

\begin{figure}
    \centering
    \includegraphics{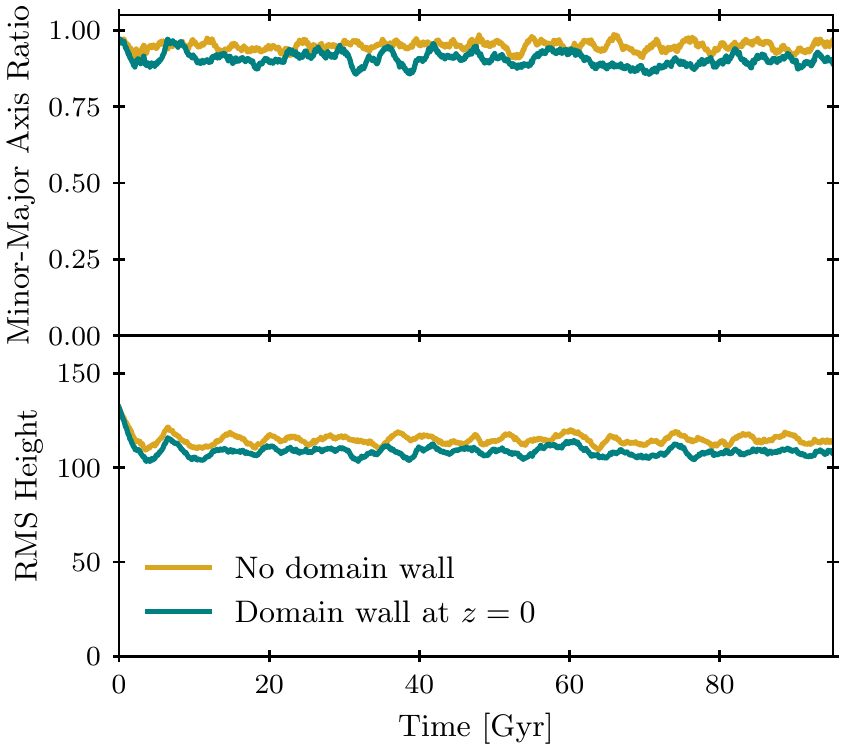}
    \caption{Two conventional planarity metrics: minor-major axis ratio (\textit{upper}) and RMS height (\textit{lower}) computed for all satellites within 400~kpc at each snapshot in the simulations with and without domain walls. By these metrics the satellites in the domain wall simulation appear only marginally more planar.}
    \label{F:Planarity}
\end{figure}

\begin{figure*}
    \centering
    \includegraphics{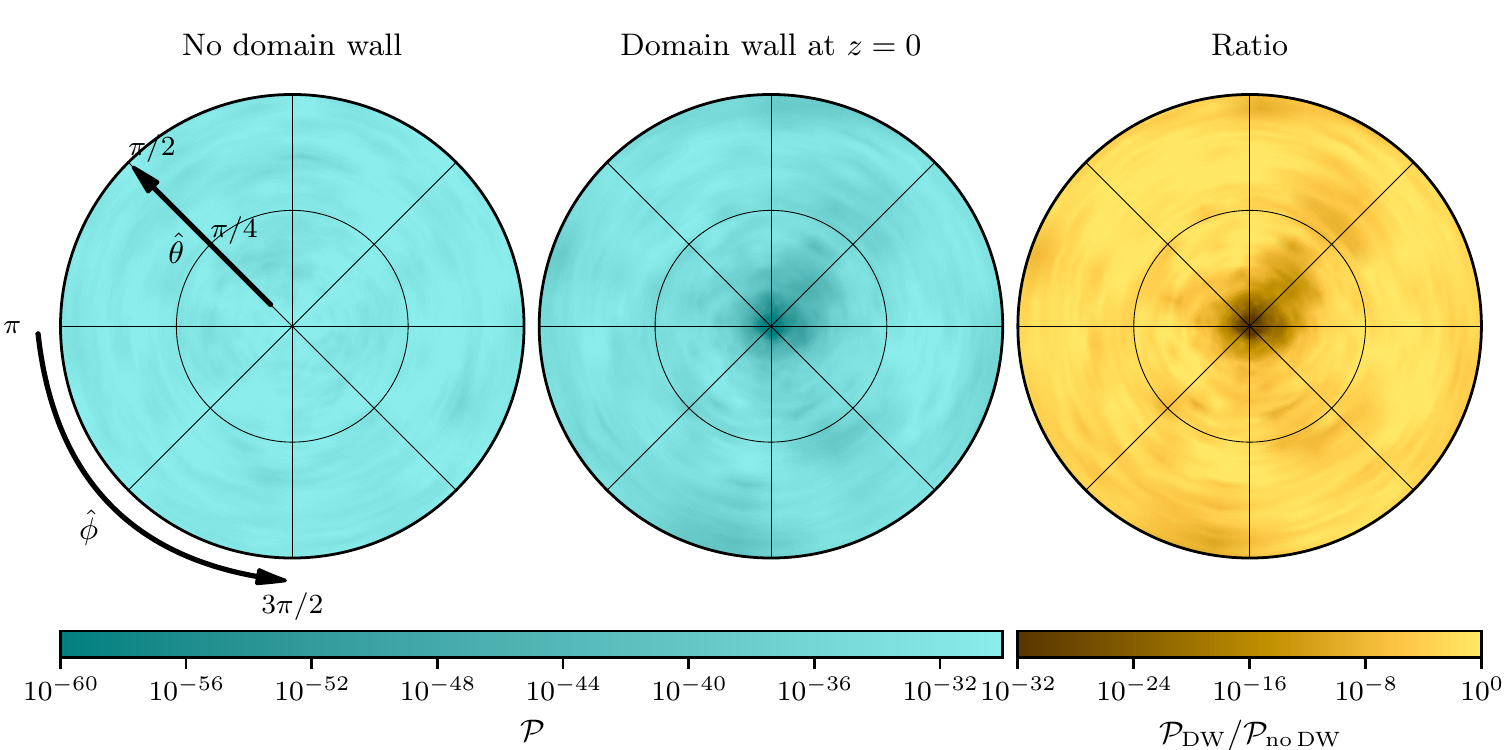}
    \caption{As a function of polar direction $(\hat{\theta}, \hat{\phi})$, the probability $\mathcal{P}$ of observing the distribution of polar angles $\theta$ of the simulated satellites (evaluated at the end of each simulation), under the assumption of isotropy. \textit{Left and middle:} $\mathcal{P}$ in the simulation without a domain wall and with a domain wall respectively. \textit{Right:} the ratio of $\mathcal{P}$ from the two simulations. When the spherical coordinate system is aligned with its polar axis perpendicular to the domain wall, the configuration of satellite positions is revealed as being highly unlikely to arise under the assumption of isotropy.}
    \label{F:IsoProb}
\end{figure*}

\section{Results}
\label{S:Results}

\subsection{Simulated Satellite Distributions}

The upper panels of Fig.~\ref{F:SatPositions} plot the positions of the satellites in the $x-z$ plane at the ends of the simulations with and without a domain wall. The effect of the domain wall is instantly clear in these panels: whereas the satellites are isotropically distributed in the simulation without a domain wall, in the presence of one there is a clear planar substructure centred around $z=0$.

This substructure is even more clear in the lower panels of Fig.~\ref{F:SatPositions}, which show the trajectories of a random subsample of 500 satellites over the course of the simulations. In the domain wall case, there is a significant subpopulation of `on-plane' satellites whose orbits remain close to and oscillate about the $z=0$ plane throughout the simulation. Meanwhile, the `off-plane' satellites exhibit more conventional near-Keplerian orbits around the host, albeit `kinked' by the domain wall attraction at each passage through $z=0$. Two orbits demonstrating these contrasting behaviours have been highlighted in the lower right panel of Fig.~\ref{F:SatPositions}, as have their counterparts in the lower left panel (i.e. the two satellites with identical initial conditions in the simulation without a domain wall).

Thus, Fig.~\ref{F:SatPositions} demonstrates that the presence of a scalar domain wall passing through a host galaxy manifests both spatially and kinematically and as a planar substructure in the orbits of the satellite galaxies. The observation of such substructure in the satellites of the Local Group or elsewhere could therefore hint at the presence of such a scalar domain wall, and it is thus worth exploring which observational diagnostics and statistics would capture this effect, particularly given that only a subset of the satellite population is confined to the vicinity of the plane. This is the subject of the following subsection.

\subsection{Observable Diagnostics}

Let us first address the kinematic clustering. In reality, we cannot observe the past trajectories of satellite galaxies, and must instead make do with instantaneous snapshots of positions and velocities. Thus, a diagram like the lower right panel of Fig.~\ref{F:SatPositions} is observationally inaccessible. In the literature, many works have quantified kinematic clustering of Milky Way satellites by considering the locations of their orbital poles on the celestial sphere. The upper panels of Fig.~\ref{F:OrbitalPoles} plot the instantaneous orbital poles of the satellites in the final snapshots of both simulations, as obtained from
\begin{equation}
    \hat{\mathbf{n}} = \frac{\mathbf{x} \times \mathbf{v}}{|\mathbf{x}||\mathbf{v}|}.
\end{equation}

Given the clustering of orbits about the domain wall mid-plane observed in Fig.~\ref{F:SatPositions}, one might expect to see a clustering of orbital poles around the north and south poles (i.e. the directions perpendicular to the plane). However, no such clustering is immediately apparent in the corresponding (upper right) panel of Fig.~\ref{F:OrbitalPoles}. Instead, the orbital poles appear no less isotropic than their counterparts in the simulation without a domain wall. The reason for this lack of clustering is the oscillation of the on-plane satellites about $z=0$, which means that the orbital poles precess around the north and south poles rather than remaining fixed there as in the case of an orbit entirely confined to the plane. 

The lower panels of Fig.~\ref{F:OrbitalPoles} show the orbital poles of the satellites averaged over the whole simulation. In the simulation without a domain wall, the spherical symmetry of the system means that angular momenta are conserved, so that $\hat{n}$ remains constant and time-averaged (lower left) and instantaneous (upper left) panels are identical. In the panel depicting the time-averaged poles from the domain wall simulation (lower right), there is now a clear clustering of orbital poles at the north and south poles, because the precession has been averaged away.

Unfortunately, like the trajectories plotted in the lower panels of Fig.~\ref{F:SatPositions}, these time-averaged orbital poles are not observationally accessible, because we can only observe instantaneous velocities. It therefore appears to be the case that the \emph{kinematic} clustering due to the domain wall is not an observable signature. However, it is worth noting that a more realistic simulation (e.g. incorporating dissipative processes) might well predict a clearer signal; this will be discussed in greater length in Sec.~\ref{S:Conclusion}.

Turning to the spatial clustering, we first consider two planarity metrics that have been widely used in the literature: the root-mean-square (RMS) height of satellites above the $z=0$ plane, and the minor-major axis ratio of the system. The latter metric is given by the ratio of the smallest and largest eigenvalues of the inertia tensor
\begin{equation}
    I_{ij} = \sum_k \left(|\mathbf{x}^{(k)}|^2\delta_{ij} - x^{(k)}_i x^{(k)}_j\right),
\end{equation}
where $\mathbf{x}^{(k)}$ is the position of satellite $k$. We calculate both metrics at all snapshots of both simulations, at each instance only including the satellites within 400~kpc of the host.

The two metrics as a function of time are plotted in the two panels of Fig.~\ref{F:Planarity}. Both metrics are reduced in the domain wall simulation at all times, but this is only a marginal decrease despite the clearly visible planar substructure of satellites in Fig.~\ref{F:SatPositions}. The reason for this modest change is that the on-plane satellites are in the minority, so these metrics are both dominated by the near-isotropic off-plane majority. It is worth noting that, as will be discussed in Sec.~\ref{S:Conclusion}, different domain wall parameter choices lead to slightly different results, but these differences are small enough that the quantitative conclusions are substantially the same. 

Given the failure of these conventional metrics to appreciably detect the domain wall signature, it is worth searching to find an alternative metric that gives a more unequivocal signal. One candidate can be found by reasoning probabilistically: it is clear from the upper right panel of Fig.~\ref{F:SatPositions} that the satellites in the domain wall simulation are not oriented isotropically, and the significance of this anisotropy can be quantified by calculating the probability of the observed angular distribution of satellites, \emph{given the assumption of isotropy}. Such an approach follows the philosophy espoused by Ref.~\cite{Pawlowski2017}, who argue that the most helpful way to analyse simulated satellite populations and place any planar configurations therein on a firm statistical footing is to calculate statistics based on the null hypothesis of isotropy.

We perform this calculation as follows. At the final snapshot in both simulations, we place all satellites into 21 bins (labelled 1, 2, ..., 21) in polar angle $\theta$ (an odd number is useful here as the central bin is then centred around the mid-plane). The bin edges $\theta_0, \theta_1, ..., \theta_{21}$ are equally spaced, and $\theta_0=0$, $\theta_{21}=\pi$. For a set of points chosen randomly on a spherical surface, the polar angles $\theta$ are distributed as $\theta \sim \frac{1}{2}\sin\theta$, so the probability of a point falling in bin $i$ is
\begin{equation}
\label{E:BinProb}
    P_i = \int_{\theta_{i-1}}^{\theta_i} \frac{1}{2}\sin\theta d\theta = \frac{1}{2}\left(\cos\theta_{i-1} - \cos\theta_{i}\right).
\end{equation}
Given $N$ such points, the probability of observing a particular set of bin occupancies $\{N_i\}$ $(\sum N_i = N)$ is then obtained from a multinomial distribution with $N$ trials and 21 outcomes,
\begin{equation}
\label{E:Multinomial}
    \mathcal{P}(\{N_i\}) = \frac{N!}{N_1! N_2! \ldots N_{21}!}P_1^{N_1}P_2^{N_2} \ldots P_{21}^{N_{21}}.
\end{equation}

We calculate this probability in the final snapshot of the simulations, and to investigate the sensitivity to choice of polar axis direction $(\hat{\theta}, \hat{\phi})$, we construct a map of $\mathcal{P}$ over the $(\hat{\theta}, \hat{\phi})$ range. These mapped probabilities are plotted in Fig.~\ref{F:IsoProb}, for the simulation without (left panel) and with (middle panel) a domain wall, and the ratio between them (right panel).

The calculated probabilities are everywhere rather small in absolute terms. This is to be expected: when $N$ is of an appreciable size, the probability of observing a \emph{specific} set of bin occupancies $\{N_i\}$ is correspondingly small. By analogy, if one performs one million coin flips, the probability of exactly 500\,000 `heads' outcomes is approximately $8 \times 10^{-4}$. Of more interest is the relative probabilities, and the variation of probability with polar axis direction.

Whereas in the simulation without a domain wall the probability is approximately constant with polar axis direction, there is a clear feature at $\hat{\theta}=0$ in the domain wall simulation, where the probability is considerably lower. In other words, if the polar axis is oriented perpendicular to the domain wall, the resulting polar angles of the satellites are such that they appear significantly less isotropic than when the polar axis points elsewhere. The reason the satellites appear near-isotropic when the polar axis is not perpendicular to the domain wall is that the on-plane satellites are consequently spread over a wide range of $\theta$ bins, rather than being concentrated in the central few bins. Considering the relative probabilities (right panel), at $\hat{\theta}=0$, the observed distribution of satellite angles in the domain wall simulation is approximately $10^{32}$ times less likely to be generated from an isotropic distribution than the distribution of satellite angles from the simulation without a domain wall. This is a clear signature of a plane of satellites.

\section{Discussion \& Conclusion}
\label{S:Conclusion}

In this contribution, we have considered a novel explanation for the observed `planes of satellites' in the Local Group and beyond: domain walls arising in theories with symmetry-breaking scalar fields coupled to matter. In comparison to domain walls that do not couple to matter, these domain walls are more numerous in the late Universe, and are `pinned' to structures of the cosmic web. To investigate this idea, we have set up and run a pair of simple simulations (one with a domain wall, one without) in which a large number of massless point particles represent satellites moving under the combined influence of a Milky Way-like host galaxy and a scalar domain wall. This work serves simply as a proof of concept, and we reserve a more sophisticated treatment to future investigations, as discussed below.

The key result of this work is encapsulated in Fig.~\ref{F:SatPositions}: in the presence of a domain wall, there is a significant planar substructure in the distribution of satellites, both spatially and kinematically. In particular, there is a subset of `on-plane' satellites whose orbits are confined to the region close to the domain wall. As they revolve around the host galaxy, they perform vertical oscillations about the domain wall mid-plane. The remaining `off-plane' satellites are almost isotropically distributed and exhibit conventional psuedo-Keplerian orbits with one key difference: a kink in the trajectory at each domain wall passage.

The remainder of this work considered several approaches to quantifying this planar substructure. Various approaches have been tried in the literature for both the kinematic and spatial clustering. For the kinematic clustering, we plotted the directions of the satellite orbital poles at the ends of both simulations. Despite the confinement of a subset of satellites to the region close to the domain wall, there is no clear difference in the orbital pole distributions in the two simulations. This is because the vertical oscillations of the on-plane satellites lead to a precession of their orbital poles, so that at any given instant there is no obvious clustering of poles on the celestial sphere. Regarding the spatial clustering, we considered two `planarity' metrics widely used in the literature: the RMS height of satellites above the plane and the minor-major axis ratio of the inertia tensor. Both metrics are only very slightly reduced by the presence of the domain wall, because the metrics are dominated by the near-isotropic off-plane satellites, which form a majority in this case.

We find that an alternative metric is altogether more discriminating: the probability of observing the distribution of satellite polar angles, assuming isotropy. If the satellites are binned in polar angle, then this is obtained simply from a multinomial distribution (Eq.~\ref{E:Multinomial}). Calculating this probability for the two simulations and taking the ratio clearly demonstrates that the satellites in the domain wall simulation are very unlikely to have been drawn from an isotropic distribution. Given a real set of observed satellites, comparing the calculated probability for their angular positions relative to a random isotropic template sample of equal size will thus reveal any such planar substructures.

It is worth noting that the results shown here correspond only to a single point in the domain wall parameter space, and it will be interesting in future to explore it more comprehensively. As a validation test, we have performed preliminary experiments in varying the domain wall thickness and characteristic acceleration, and found that our results are qualitatively robust, although the precise magnitudes of the resulting satellite planes and the discriminating power of various observational diagnostics can vary. As one might expect, increasing the acceleration parameter leads to more distinct planes and vice versa, while increasing the wall thickness leads to more populated satellite planes and vice versa (although the thinner domain walls might have proved more capable of capturing satellites had dissipative physics been included). 

It is interesting that the domain wall led to distinct subpopulations of on-plane and off-plane satellites, and that this in turn led to the `conventional' planarity metrics largely failing to detect a plane of satellites. The satellite galaxies of Andromeda have been observed to be similarly bimodal, with an on-plane subset very tightly confined to a plane and a near-isotropic off-plane subset \citep{Ibata2013, Conn2013}. Perhaps as a consequence, the statistical significance of Andromeda's plane of satellites has been refuted to a greater extent than that of the Milky Way (e.g.~\cite{Cautun2015}). It will be worth reconsidering the satellites of Andromeda in light of our present findings.

That being said, our treatment is merely a proof of concept, and adopts various unrealistic simplifications which perhaps render a quantitative comparison to observations premature. First, our simulation incorporates minimal physics, but there are relevant factors such as dissipative processes that could have an appreciable effect, perhaps to our benefit: under the present treatment, on-plane satellites are able to oscillate vertically about the mid-plane without any damping, in some cases completing many such oscillations on each revolution about the host galaxy. In reality, dynamical friction and other such dissipative effects would work over time to deplete this vertical energy and confine the satellite's orbit to the mid-plane and end the precession of the orbital pole. Consequently, a clustering of orbital poles should re-emerge, and the various spatial metrics should give a stronger signal. 

Second, our isotropic initial conditions are rather simplistic. A more sophisticated approach incorporating a preferred direction for satellite infall could work both ways, i.e. to decrease or increase the significance of the satellite plane. One the one hand, a preferred infall direction might give the satellite population an intrinsic anisotropy, leading to a false signal in a probabilistic treatment which assumes isotropy as a null hypothesis such as Eq.~\ref{E:Multinomial}. On the other hand, a preferred infall direction could account for an observed phenomenon hitherto unexplained by our model: under our treatment, an on-plane satellite is equally likely to revolve clockwise or anti-clockwise about the host galaxy (e.g., the lower right panel of Fig.~\ref{F:OrbitalPoles} shows clusters at both the north and south poles), whereas in reality the observed on-plane satellites share their sense of co-rotation. It can be imagined that a preferred direction of initial infall would subsequently mean a preferred sense of rotation.

Third, our simulated satellite populations are much larger than realistic satellite numbers, which might mean that the statistical treatment we proposed (Eq.~\ref{E:Multinomial}) is less effective. We have experimented with taking random subsamples of our simulated satellites and we find that the question of whether a given symmetron realisation leads to a statistically significant satellite plane in the low $N$ limit depends sensitively on the `on-plane' fraction of satellites, which we perhaps under-predict in our dissipationless simulations.

Finally, our treatment of the scalar field and domain wall could also be made more sophisticated. We assumed a stable, static, flat domain wall, but an unstable, short-lived or highly curved domain wall might prevent the formation of a plane of satellites, although the pinning phenomenon gives us a measure of confidence that the satellites will see a domain wall that is at least locally flat and stable. Moreover, a key aspect of e.g. symmetron theories is the screening mechanism, under which sufficiently dense objects neither source nor couple to an external fifth force. We assumed all of our satellites feel the fifth force, but if all of the satellites are screened then no plane could form. On the other hand, if only some of the satellites are screened, there would be a correlation between satellite density and plane membership: a clear signal that a plane of satellites has arisen due to a domain wall with a screenable fifth force. 

We propose to address all of these outstanding issues in a future work, in which we will run full N-body simulations incorporating a symmetron scalar field and fifth force, allowing for the organic formation of domain walls and resultant planes of satellites. Running this simulation with initial conditions `constrained' to resemble the Local Group would render it even more useful: the satellite populations extracted from this simulation will be ready for a direct, quantitative comparison to the observed planes of satellites in the local Universe.

\acknowledgments

We thank Bradley March for useful comments. APN and CB are supported by a Research Leadership Award from the Leverhulme Trust. We are grateful for access to the University of Nottingham's Augusta HPC service.

\bibliographystyle{JHEP}
\bibliography{library}

\end{document}